\documentclass{ws-procs961x669}            

\begin{document}
\title{Mass structure and pressure forces inside the nucleon}

\author{C. Lorc\'e}

\address{CPHT, CNRS, Ecole Polytechnique, Institut Polytechnique de Paris,\\ Route de Saclay, 91128 Palaiseau, France\\
E-mail: cedric.lorce@polytechnique.edu}

\begin{abstract}
We summarize recent works on the question of the nucleon mass decomposition and the 2D relativistic distribution of pressure forces on the light front. All these mechanical properties are encoded in the energy-momentum tensor of the system which can be constrained using various types of high-energy lepton-nucleon scatterings. Some further developments for targets with spin $>1/2$ are also reported.
\end{abstract}


\bodymatter

\section{Introduction}
Elastic scattering on the nucleon provided during the last 60 years key information about how electric charge and magnetization are distributed in position space within the nucleon~\cite{Perdrisat:2006hj,Miller:2010nz}. 20 years ago, other exclusive reactions like e.g. deeply virtual Compton scattering and meson production were shown to give access to generalized parton distributions (GPDs), which are the mother distributions of both ordinary parton distributions and electromagnetic form factors~\cite{Diehl:2003ny,Belitsky:2005qn,Kumericki:2016ehc}. On top of providing tomographic pictures of the internal structure of the nucleon, GPDs also give access to the gravitational form factors (GFFs) which characterize the energy-momentum tensor (EMT)~\cite{Ji:1996ek}. Just like the Fourier transform of electromagnetic form factors can be interpreted in terms of spatial distribution of electric charge and magnetization, the Fourier transform of GFFs can be interpreted in terms of spatial distribution of energy, momentum and pressure forces~\cite{Polyakov:2002yz,Polyakov:2018zvc}. 

We present in this contribution a short summary of some recent developments providing a new look on the nucleon internal structure.

\section{Mass decomposition and balance equations}
Because of Poincar\'e symmetry, one can write in general for the matrix element of the EMT $T^{\mu\nu}(0)$ as follows
\begin{equation}\label{Poincare}
\langle P|T^{\mu\nu}(0)|P\rangle=2P^\mu P^\nu
\end{equation}
for a spin-$1/2$ state with relativistic normalization $\langle p'|p\rangle=2p^0(2\pi)^3\delta^{(3)}(\vec p'-\vec p)$. It is then clear that the mass $M$ of the nucleon can be expressed in an explicitly covariant way in terms of the trace of the EMT~\cite{Shifman:1978zn,Jaffe:1989jz}
\begin{equation}
\langle P|T^{\mu}_{\phantom{\mu}\mu}(0)|P\rangle=2M^2.
\end{equation}
The trace of the QCD EMT tensor being given by
\begin{equation}
T^{\mu}_{\phantom{\mu}\mu}=\frac{\beta(g)}{2g}\,G^2+(1+\gamma_m)\,\overline\psi m\psi,    
\end{equation}
it is tempting to interpret $\langle P|\frac{\beta(g)}{2g}\,G^2|P\rangle$ and $\langle P|(1+\gamma_m)\,\overline\psi m\psi|P\rangle$ as the gluon and quark contributions to the nucleon mass, respectively. This is however incorrect since the partial (quark or gluon) EMT is not conserved and reads in general~\cite{Ji:1994av,Lorce:2017xzd}
\begin{equation}
\langle P|T^{\mu\nu}_{q,G}(0)|P\rangle=2P^\mu P^\nu A_{q,G}(0)+2M^2\eta^{\mu\nu}\bar C_{q,G}(0),
\end{equation}
where $A_{q,G}(t)$ and $\bar C_{q,G}(t)$ are GFFs depending on the squared four-momentum transfer $t=\Delta^2=(p'-p)^2$. Unlike electromagnetic form factors, GFFs also depend on the renormalization scale and scheme. The extra term accounts for the non-conservation of the partial EMT. From a more physical point of view, it is related to the pressure-volume work exerted by the quark and gluon subsystems. The nucleon being a stable object, the total pressure-volume work has to vanish and therefore disappears once the EMT is summed over quark and gluon contributions like in Eq.~\eqref{Poincare}. Paying attention to properly distinguish contributions to energy and pressure-volume work, one arrives at a proper mass decomposition and a balance equation~\cite{Lorce:2017xzd}
\begin{equation}
M=U_q+U_G,\qquad W_q+W_G=0
\end{equation}
with $U_{q,G}=\left[A_{q,G}(0)+\bar C_{q,G}\right]M$ and $W_{q,G}=-\bar C_{q,G}M$. Using current phenomenological estimates~\cite{Gao:2015aax}, one finds that $U_q\approx 0.44 M$ and $W_q\approx0.11 M$. Contrary to what is sometimes claimed in the literature, the large value of $\langle P|\frac{\beta(g)}{2g}\,G^2|P\rangle/2M=U_G-3W_G\approx 0.89 M$ does not indicate that gluons are responsible for most of the nucleon mass, but comes from the fact that the gluon pressure-volume work is large (reflecting the relativistic nature of the nucleon) and negative (i.e. attractive).

\section{Relativistic 2D distributions of pressure forces}
By analogy with electromagnetic form factors, one can interpret the Fourier transform of GFFs in the Breit frame $\vec P=(\vec p^{\,\prime}+\vec p)/2=\vec 0$ in terms of 3D spatial distribution of mass, momentum and pressure forces~\cite{Polyakov:2002yz}. Distributions defined in this way are however known to be plagued by relativistic corrections~\cite{Burkardt:2000za}. The latter can be avoided by considering instead transverse 2D spatial distributions within the light-front formalism. The ones associated with the EMT are defined in the symmetric Drell-Yan frame $\vec P_\perp=\vec 0_\perp$, $\Delta^+=0$ as~\cite{Lorce:2017wkb,Lorce:2018egm}
\begin{equation}\label{EMTdist}
\langle T^{\mu\nu}_{q,G}\rangle(\vec b_\perp)=\int\frac{\textrm{d}^2\Delta_\perp}{(2\pi)^2}\,e^{-i\vec\Delta_\perp\cdot\vec b_\perp}\,\frac{1}{2P^+}\langle P^+,\tfrac{\vec\Delta_\perp}{2}|T^{\mu\nu}_{q,G}(0) |P^+, -\tfrac{\vec\Delta_\perp}{2}\rangle
\end{equation}
with $a^+=(a^0+a^3)/\sqrt{2}$. The $T^{++}$ component plays the role of gallilean mass in the transverse plane and has been studied in Refs.~\citenum{Abidin:2008sb,Lorce:2018zpf}. The longitudinal orbital angular momentum of quarks and gluons can be obtained from the $T^{+i}$ components~\cite{Lorce:2017wkb} and the transverse $T^{ij}$ components can be interpreted in terms of 2D pressure forces~\cite{Lorce:2018egm}. Remarkably, in the last two cases the relativistic 2D distributions appear to coincide with the projection of the 3D distributions defined in the Breit frame onto the transverse plane.

One can write in the transverse plane~\cite{Lorce:2018egm}
\begin{equation}
\langle T^{ij}_{q,G}\rangle(\vec b_\perp)=\sigma(b)\,\delta^{ij}_\perp+\Pi(b)\left(\frac{b^i_\perp b^j_\perp}{b^2}-\delta^{ij}_\perp\right),
\end{equation}
where $b=|\vec b_\perp|$, $\sigma(b)$ represents 2D isotropic pressure and $\Pi(b)$ represents 2D pressure anisotropy. In non-relativistic systems, pressure anisotropy is confined to a very thin region at the boundary and described by a surface tension. In relativistic systems like the nucleon and compact stars, pressure anisotropy extends over a larger region in the bulk. Using a multipole parametrization for the nucleon GFFs~\cite{Lorce:2018egm}, it appeared that the quark contribution to the EMT is mostly repulsive and short range, while the gluon contribution is mostly attractive and long range. This configuration ensures naturally the mechanical stability of the nucleon as a whole.

\section{Higher-spin targets}
Studies of the EMT mostly focused on the nucleon, or more generally spin-$1/2$ targets. Higher-spin targets are however of deep interest in both hadronic and nuclear physics. Recently, a detailed study of the spin-$1$ case has been conducted in Ref.\citenum{Cosyn:2019aio}. It indicated in particular that higher-spin targets simply involve additional contributions associated with higher spin-multipoles, suggesting an alternative approach based on a covariant multipole expansion (still under development). Poincar\'e symmetry has been used to constrain some of the GFFs and to derive the Ji relation for arbitrary spin targets~\cite{Cotogno:2019xcl,Lorce:2019sbq}.

\section{Conclusion}
We presented a short summary of recent developments about the energy-momentum tensor of hadrons. A tomography of the origin of mass and spin along with pressure forces is now possible. Stability conditions may provide new constraints on the observables and hints about the mechanism of confinement. In the coming years, further constraints on gravitational form factors using both exclusive high-energy experiments and Lattice QCD are expected, bringing our understanding of the hadron internal structure to a whole new and exciting level.

\section*{Acknowledgements}
Some of the works presented here have been supported by the Agence Nationale de la Recherche (ANR-16-CE31-0019 and ANR-18-ERC1-0002), the P2IO LabEx (ANR-10-LABX-0038) in the framework ``Investissements d'Avenir'' (ANR-11-IDEX-0003-01) managed by the Agence Nationale de la Recherche, the CEA-Enhanced Eurotalents Program co-funded by FP7 Marie Sklodowska-Curie COFUND Program (No. 600382), the U.S. Department of Energy, Office of Science, Office of Nuclear Physics (No. DE-AC02-06CH11357), and an LDRD initiative at Argonne National Laboratory (No. 2017-058-N0).



\begin{thebibliography}{99}

\bibitem{Perdrisat:2006hj} 
  C.~F.~Perdrisat, V.~Punjabi and M.~Vanderhaeghen,
  Nucleon Electromagnetic Form Factors,
  {\em Prog.\ Part.\ Nucl.\ Phys.\ }{\bf 59}, 694 (2007).
  
\bibitem{Miller:2010nz} 
  G.~A.~Miller,
  Transverse Charge Densities,
  {\em Ann.\ Rev.\ Nucl.\ Part.\ Sci.\ }{\bf 60}, 1 (2010).

\bibitem{Diehl:2003ny} 
  M.~Diehl,
  Generalized parton distributions,
  {\em Phys.\ Rept.\  }{\bf 388}, 41 (2003).
  
\bibitem{Belitsky:2005qn} 
  A.~V.~Belitsky and A.~V.~Radyushkin,
  Unraveling hadron structure with generalized parton distributions,
  {\em Phys.\ Rept.\  }{\bf 418}, 1 (2005).
  
\bibitem{Kumericki:2016ehc} 
  K.~Kumericki, S.~Liuti and H.~Moutarde,
  GPD phenomenology and DVCS fitting : Entering the high-precision era,
  {\em Eur.\ Phys.\ J.\ A} {\bf 52}, no. 6, 157 (2016).

\bibitem{Ji:1996ek} 
  X.~D.~Ji,
  Gauge-Invariant Decomposition of Nucleon Spin,
  {\em Phys.\ Rev.\ Lett.\  }{\bf 78}, 610 (1997).

\bibitem{Polyakov:2002yz} 
  M.~V.~Polyakov,
  Generalized parton distributions and strong forces inside nucleons and nuclei,
  {\em Phys.\ Lett.\ B} {\bf 555}, 57 (2003).

\bibitem{Polyakov:2018zvc} 
  M.~V.~Polyakov and P.~Schweitzer,
  Forces inside hadrons: pressure, surface tension, mechanical radius, and all that,
  {\em Int.\ J.\ Mod.\ Phys.\ A} {\bf 33}, no. 26, 1830025 (2018).

\bibitem{Shifman:1978zn} 
  M.~A.~Shifman, A.~I.~Vainshtein and V.~I.~Zakharov,
  Remarks on Higgs Boson Interactions with Nucleons,
  {\em Phys.\ Lett.\  }{\bf 78B}, 443 (1978).

\bibitem{Jaffe:1989jz} 
  R.~L.~Jaffe and A.~Manohar,
  The G(1) Problem: Fact and Fantasy on the Spin of the Proton,
  {\em Nucl.\ Phys.\ B} {\bf 337}, 509 (1990).

\bibitem{Ji:1994av} 
  X.~D.~Ji,
  A QCD analysis of the mass structure of the nucleon,
  {\em Phys.\ Rev.\ Lett.\  }{\bf 74}, 1071 (1995).

\bibitem{Lorce:2017xzd} 
  C.~Lorc\'e,
  On the hadron mass decomposition,
  {\em Eur.\ Phys.\ J.\ C} {\bf 78}, no. 2, 120 (2018).

\bibitem{Gao:2015aax} 
  H.~Gao, T.~Liu, C.~Peng, Z.~Ye and Z.~Zhao,
  Proton Remains Puzzling,
  {\em The Universe} {\bf 3}, no. 2, 18 (2015).

\bibitem{Burkardt:2000za} 
  M.~Burkardt,
  Impact parameter dependent parton distributions and off forward parton distributions for zeta $\to$ 0,
  {\em Phys.\ Rev.\ D} {\bf 62}, 071503 (2000)
  Erratum: [{\em Phys.\ Rev.\ D} {\bf 66}, 119903 (2002)].

\bibitem{Lorce:2017wkb} 
  C.~Lorc\'e, L.~Mantovani and B.~Pasquini,
  Spatial distribution of angular momentum inside the nucleon,
  {\em Phys.\ Lett.\ B} {\bf 776}, 38 (2018).

\bibitem{Lorce:2018egm} 
  C.~Lorc\'e, H.~Moutarde and A.~P.~Trawi\'nski,
  Revisiting the mechanical properties of the nucleon,
  {\em Eur.\ Phys.\ J.\ C} {\bf 79}, no. 1, 89 (2019).

\bibitem{Abidin:2008sb} 
  Z.~Abidin and C.~E.~Carlson,
  Hadronic Momentum Densities in the Transverse Plane,
  {\em Phys.\ Rev.\ D} {\bf 78}, 071502 (2008).

\bibitem{Lorce:2018zpf} 
  C.~Lorc\'e,
  The relativistic center of mass in field theory with spin,
  {\em Eur.\ Phys.\ J.\ C} {\bf 78}, no. 9, 785 (2018).

\bibitem{Cosyn:2019aio} 
  W.~Cosyn, S.~Cotogno, A.~Freese and C.~Lorc\'e,
  The energy-momentum tensor of spin-1 hadrons: formalism,
  {\em Eur.\ Phys.\ J.\ C} {\bf 79}, no. 6, 476 (2019).

\bibitem{Cotogno:2019xcl} 
  S.~Cotogno, C.~Lorc\'e and P.~Lowdon,
  Poincaré constraints on the gravitational form factors for massive states with arbitrary spin,
  {\em Phys.\ Rev.\ D} {\bf 100}, no. 4, 045003 (2019).

\bibitem{Lorce:2019sbq} 
  C.~Lorc\'e and P.~Lowdon,
  Universality of the Poincaré gravitational form factor constraints,
  arXiv:1908.02567 [hep-th].
  
  

\end{thebibliography}
\end{document}